\documentclass[twocolumn,nofootinbib]{revtex4}
\usepackage{graphicx}
\usepackage{latexsym}
\usepackage[colorlinks, linkcolor=blue, citecolor=blue, urlcolor=blue]{hyperref}

\begin{document}

\title{Constraints on Scalar Spectral Index from Latest Observational Measurements}

\author{Hong Li${}^{a,b}$}
\author{Jun-Qing Xia${}^{a}$}
\author{Xinmin Zhang${}^{c}$}

\affiliation{${}^a$Key Laboratory of Particle Astrophysics, Institute of High Energy Physics, Chinese Academy of Science, P. O. Box 918-3, Beijing 100049, P. R. China}
\affiliation{${}^b$National Astronomical Observatories, Chinese Academy of Sciences, Beijing 100012, P. R. China}
\affiliation{${}^c$Theoretical Physics Division, Institute of High Energy Physics, Chinese Academy of Science, P. O. Box 918-4, Beijing 100049, P. R. China}

\date{\today}

\begin{abstract}

Recently, the nine-year data release of the Wilkinson Microwave Anisotropy Probe (WMAP9) found that the inflationary models with the scalar spectral index $n_s \geq 1$ are excluded at about $5\,\sigma$ confidence level. In this paper, we set the new limits on the scalar spectral index in different cosmological models combining the WMAP9 data with the small-scale cosmic microwave background measurement from the South Pole Telescope, baryon acoustic oscillation data, Hubble Telescope measurements of the Hubble constant, and supernovae luminosity distance data. In most of extended cosmological models, e.g. with a dark energy equation of state, the constraints on $n_s$ do not change significantly. The Harrison-Zel'dovich-Peebles (HZ) scale invariant spectrum is still disfavored at more than $4\,\sigma$ confidence level. However, when considering the model with a number of relativistic species $N_{\rm eff}$, we obtain the limit on the spectral index of $n_s=0.980\pm0.011$ ($1\,\sigma$), due to the strong degeneracy between $n_s$ and $N_{\rm eff}$. The HZ spectrum now is still consistent with the current data at $95\%$ confidence level.

\end{abstract}


\maketitle


\section{Introduction}\label{Int}

Inflation, the most attractive paradigm in the very early universe, has successfully resolved many problems existing in the hot big bang cosmology, such as the flatness, horizon, monopole problem, and so forth \cite{infpaper}. Its quantum fluctuations turn out to be the primordial density fluctuations which seed the observed large scale structures (LSS) and the anisotropies of cosmic microwave background (CMB). To distinguish various inflationary models, the spectral index of the power spectrum of primordial curvature perturbations is one of the most important variables.

With the accumulation of observational data from CMB, LSS and Type Ia Supernovae observations (SN) and the improvements of the data quality, the cosmological observations play a crucial role in our understanding of the Universe and also in constraining the cosmological parameters \cite{Xia:2008ex,Li:2008vf,Li:2009cu,Li:2012vn}. Thus, determining the scalar spectral index $n_s$ from the observational data is a very powerful and reliable way to understand these inflationary models.

Recently, the new CMB data have been released \cite{wmap9,Bennett:2012fp}. The nine-year data release of Wilkinson Microwave Anisotropy Probe (WMAP9) has determined the cosmological parameters accurately and found that the $68\%$ C.L. constraint on the scalar spectral index of $n_s=0.9608\pm0.0080$ \cite{wmap9}, when combining with the small-scale CMB measurement from the South Pole Telescope, baryon acoustic oscillation data (BAO) and Hubble Telescope measurements of the Hubble constant (HST). Within the $\Lambda$CDM framework, the Harrison-Zel'dovich-Peebles (HZ) scale invariant spectrum ($n_s \equiv 1$) and the spectra with $n_s > 1$ are disfavored by about $5\,\sigma$ confidence level.

Although the $\Lambda$CDM model is a good candidate for interpreting the data, the evidence for various extended models should not be neglected, such as the number of relativistic species \cite{Xia:2012na,Feeney:2013wp}, the massive neutrino \cite{Zhao:2012xw,Mak:2013jia} or the equation of state of dark energy \cite{Li:2012ug}. More importantly, the degeneracies between $n_s$ and cosmological parameters introduced in these models could weaken the constraint on $n_s$ \cite{xiainf,slowroll,xia2008}.

In this paper, we explore the cosmological constraints on $n_s$ in some extended models from the latest cosmological data sets, including the WMAP9 temperature and polarization power spectra, the small-scale CMB measurement from SPT, the BAO measurements from several LSS surveys, the HST prior on the Hubble constant $H_0$ and the ``Union2.1'' compilation SN sample made by the Supernova Cosmology Project. Firstly, we consider the general inflationary model with the tensor fluctuations ($r$) in the $\Lambda$CDM framework. We then extend the $\Lambda$CDM model allowing for the dark energy models with a constant equation of state (EoS, $w$) or with a time-varying EoS ($w(z)$). Finally, we include the massive neutrino case ($\sum{m_\nu}$) or the number of relativistic species ($N_{\rm eff}$) in the $\Lambda$CDM model.

Our paper is organized as follows: In Section \ref{method} we describe the method and the latest observational data sets used in the numerical analyses; Section \ref{result} contains our main global constraints of the scalar spectral index $n_s$ in different cosmological models from the current observations. The last Section \ref{summary} is the conclusions.


\section{Method and Data}\label{method}

\subsection{Numerical Method}

We perform a global fitting of cosmological parameters using the {\tt CosmoMC} package \cite{cosmomc}, a Markov Chain Monte Carlo (MCMC) code. We assume purely adiabatic initial conditions and a flat $\Lambda$CDM Universe. The following six cosmological parameters are allowed to vary with top-hat priors: the cold dark matter energy density parameter $\Omega_ch^2 \in [0.01, 0.99]$, the baryon energy density parameter $\Omega_bh^2 \in [0.005, 0.1]$, the scalar spectral index $n_s \in [0.5, 1.5]$, the primordial amplitude $\ln[10^{10}A_s] \in [2.7, 4.0]$, the ratio (multiplied by 100) of the sound horizon at decoupling to the angular diameter distance to the last scattering surface $100\Theta_s \in [0.5, 10]$, and the optical depth to reionization $\tau \in [0.01, 0.8]$. The pivot scale is set at $k_{s0} = 0.05{\rm Mpc}^{-1}$. Besides these six basic cosmological parameters, we have several extra cosmological parameters in different extended cosmological models: the running of scalar spectral index $\alpha_s\equiv d\ln n_s/d\ln k \in [-0.1,0.1]$; the tensor to scalar ratio of the primordial spectrum $r \equiv A_t/A_s \in [0,2]$; the total neutrino mass fraction at the present day
\begin{equation}
f_{\nu}\equiv\frac{\Omega_{\nu}h^2}{\Omega_mh^2}=\frac{\sum{
m_{\nu}}}{93.14~\mathrm{eV}~\Omega_mh^2}\in[0,0.1]~;
\end{equation}
and the number of relativistic species $N_{\rm eff}\in[0,10]$. We also consider the dark energy model with the EoS parameters $w_0 \in [-2,0]$ and $w_1\in[-5,2]$, which is given by the parametrization \cite{cpl}
\begin{equation}
w_{de}(a) = w_{0} + w_{1}(1-a)~,
\end{equation}
where $a\equiv1/(1+z)$ is the scale factor and $w_{1}=-dw/da$ characterizes the ``running'' of EoS. The $\Lambda$CDM model has $w_0=-1$ and $w_1=0$. For the dark energy model with a constant EoS, $w_1=0$. When using the global fitting strategy to constrain the cosmological parameters, it is crucial to include dark energy perturbations \cite{xiapert}. In this paper we use the method provided in refs. \cite{xiapert,zhaopert} to treat the dark energy perturbations consistently in the whole parameter space in the numerical calculations. Therefore, the most general parameter space in the analyses is:
\begin{equation}
\{\Omega_{b}h^2, \Omega_{c}h^2, \Theta_{s}, \tau, n_s, A_s,\alpha_s,r,w_{0}, w_{1}, f_{\nu}, N_{\rm eff}\}~.
\end{equation}

\subsection{Current Observational Data}

In our analysis, we consider the following cosmological probes: i) power spectra of CMB temperature and polarization anisotropies; ii) the baryon acoustic oscillation in the galaxy power spectra; iii) measurement of the current Hubble constant; iv) luminosity distances of type Ia supernovae.

To incorporate the WMAP9 CMB temperature and polarization power spectra, we use the routines for computing the likelihood supplied by the WMAP team \cite{wmap9}.  The WMAP9 polarization data are composed of TE/EE/BB power spectra on large scales ($2 \leq \ell \leq 23$) and TE power spectra on small scales ($24 \leq \ell \leq 800$), while the WMAP9 temperature data includes the CMB anisotropies on scales $2 \leq \ell \leq 1200$. Furthermore, we also use the recent SPT data \cite{spt}, using 47 bandpowers in the range $600 \leq \ell \leq 3000$. The likelihood is assumed to be Gaussian, and we use the published band-power window functions and covariance matrix. In order to address for foreground contributions, the SZ amplitude, the amplitude of the clustered point source contribution, and the amplitude of the Poisson distributed point source contribution, are added as nuisance parameters in the CMB data analyses.

Baryon Acoustic Oscillations provides an efficient method for measuring the expansion history by using features in the clustering of galaxies within large scale surveys as a ruler with which to measure the distance-redshift relation. It provides a particularly robust quantity to measure \cite{bao}. It measures not only the angular diameter distance, $D_A(z)$, but also the expansion rate of the universe, $H(z)$, which is powerful for studying dark energy \cite{task}. Since the current BAO data are not accurate enough for extracting the information of $D_A(z)$ and $H(z)$ separately \cite{okumura}, one can only determine an effective distance \cite{baosdss}:
\begin{equation}
D_v(z)=[(1+z)^2D_A^2(z)cz/H(z)]^{1/3}~.
\end{equation}
In this paper we use the recent BAO measurement at high redshift $z=2.3$ detected in the Ly-$\alpha$ forest of Baryon Oscillation Spectroscopic Survey (BOSS) quasars \cite{highbao}. Furthermore, we also include the BAO measurement from the 6dF Galaxy Redshift Survey (6dFGRS) at a low redshift $z = 0.106$ \cite{6dfgrs}, and the BAO measurements from the WiggleZ Survey at three redshift bins $z=0.44$, $z=0.60$ and $z=0.73$ \cite{wigglez}, the measurement of the BAO scale based on a re-analysis of the Luminous Red Galaxies (LRG) sample from Sloan Digital Sky Survey (SDSS) Data Release 7 at the median redshift $z=0.35$ \cite{sdssdr7}, and the BAO signal from BOSS CMASS DR9 data at $z=0.57$ \cite{sdssdr9}.

In our analysis, we add a Gaussian prior on the current Hubble constant given by ref. \cite{h0}; $H_0 = 73.8 \pm 2.4$ km\,s${}^{-1}$\,Mpc${}^{-1}$ (68\% C.L.). The quoted error includes both statistical and systematic errors. This measurement of $H_0$ is obtained from the magnitude-redshift relation of 240 low-z Type Ia supernovae at $z < 0.1$ by the Near Infrared Camera and Multi-Object Spectrometer (NICMOS) Camera 2 of the Hubble Space Telescope (HST). This is a significant improvement over the previous prior, $H_0 = 72 \pm 8$ km\,s${}^{-1}$\,Mpc${}^{-1}$, which is from the Hubble Key project final result. In addition, we impose a weak top-hat prior on the Hubble parameter: $H_0 \in [40, 100]$ km\,s${}^{-1}$\,Mpc${}^{-1}$.

Finally, we include data from Type Ia supernovae, which consists of luminosity distance measurements as a function of redshift, $D_L(z)$. In this paper we use the latest SN data sets from the Supernova Cosmology Project, ``Union Compilation 2.1'', which consists of 580 samples and spans the redshift range $0 \leq z\leq1.55$ \cite{union2.1}. This data set also provides the covariance matrix of data with and without systematic errors. In order to be conservative, we use the covariance matrix with systematic errors. When calculating the likelihood from SN, we marginalize over the absolute magnitude M, which is a nuisance parameter, as done in refs. \cite{SNMethod}.

\section{Numerical Results}\label{result}

In this section we present our global fitting results of the cosmological parameters determined from the latest observational data and focus on the degeneracies between $n_s$ and other extended parameters in different models.

\begin{figure}[t]
\begin{center}
\includegraphics[scale=0.45]{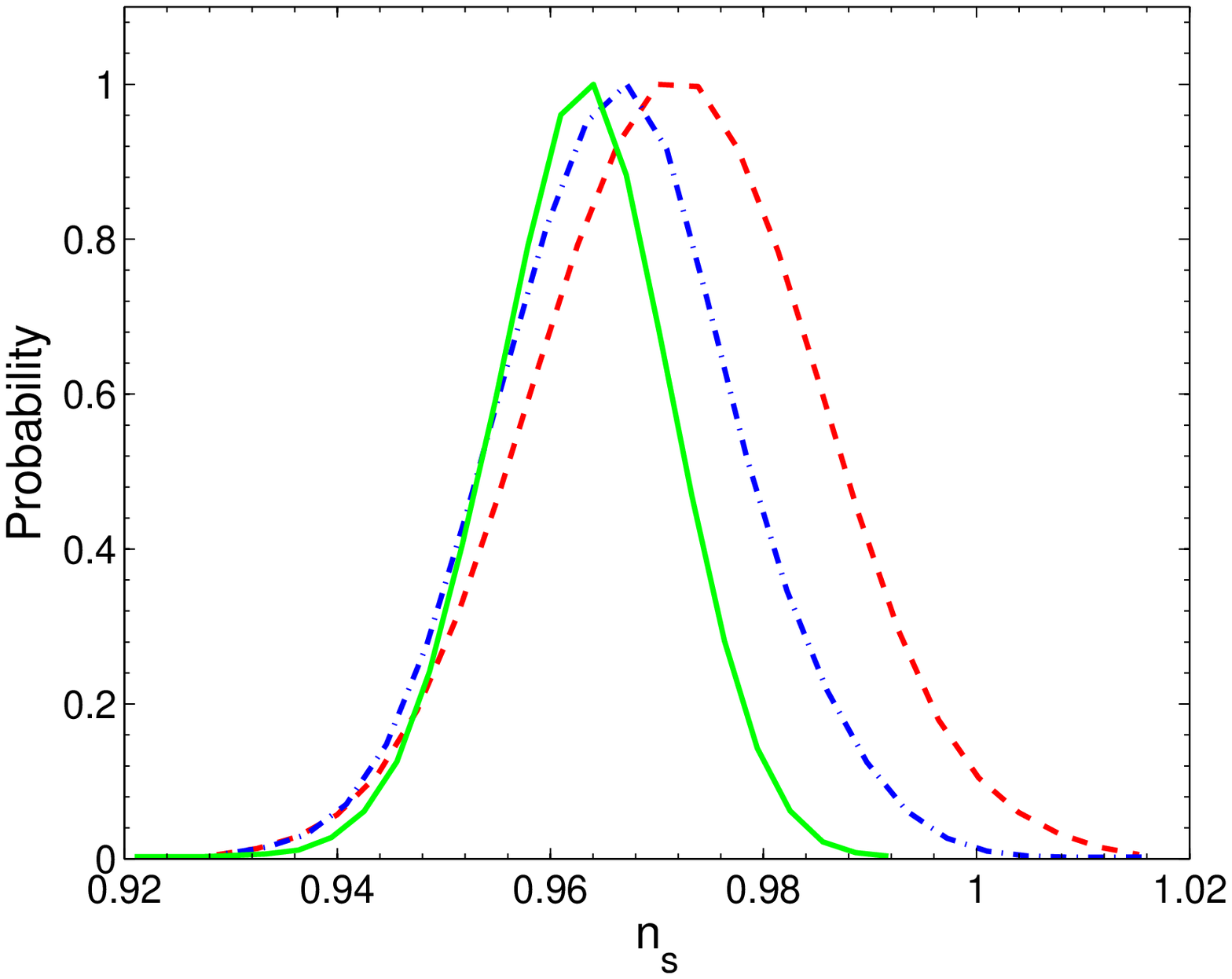}
\includegraphics[scale=0.45]{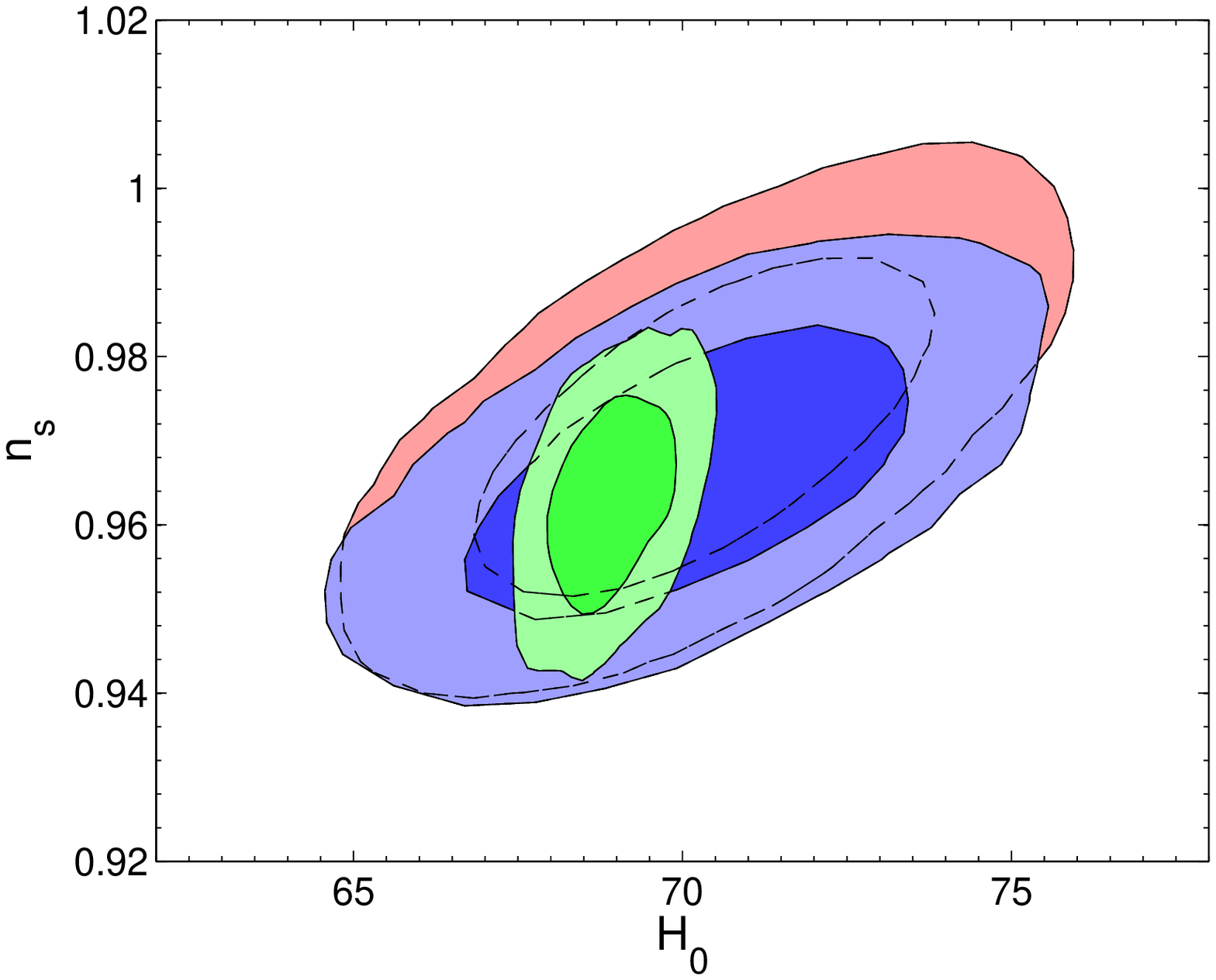}
\caption{Marginalized one-dimensional and two-dimensional likelihood (1, $2\,\sigma$ contours) constraints on the parameters $n_s$ and $H_0$ in the standard $\Lambda$CDM model from different present data combinations: WMAP9 only (red), WMAP9+SPT (blue) and All datasets (green). \label{lcdm}}
\end{center}
\end{figure}

\begin{table}
\caption{$1\,\sigma$ constraints on some cosmological parameters from different data combinations in the standard $\Lambda$CDM model.}\label{lcdmtable}
\begin{center}

\begin{tabular}{c|c|c|c}

\hline\hline
\multicolumn{4}{c}{Standard $\Lambda$CDM}\\
\hline
&WMAP9 alone&WMAP9+SPT&All Datasets\\
\hline
$n_s$&$0.972\pm0.013$&$0.966\pm0.011$&$0.963\pm0.008$\\
$H_0$&$70.34\pm2.21$&$70.20\pm2.15$&$68.90\pm0.62$\\
$100\Omega_bh^2$&$2.270\pm0.050$&$2.234\pm0.042$&$2.224\pm0.034$\\
$100\Omega_ch^2$&$11.37\pm48$&$11.41\pm0.47$&$11.70\pm0.17$\\

\hline\hline
\end{tabular}
\end{center}
\end{table}

\subsection{Standard $\Lambda$CDM model}

Firstly, we consider the standard $\Lambda$CDM model. In table \ref{lcdmtable} we show the constraints on some related cosmological parameters from three different data combinations: WMAP9 alone, WMAP9+SPT, and All datasets. In the upper panel of figure \ref{lcdm} we show the one-dimensional marginalized likelihood distributions of $n_s$ from three data combinations. Using the WMAP9 data alone, we obtain the $68\%$ constraint of $n_s=0.972\pm0.013$. The primordial spectra with $n_s \geq 1$ are only excluded at $2\,\sigma$ confidence level. When we include the small-scale SPT measurement, the constraint on $n_s$ becomes slightly tighter, $n_s=0.966 \pm 0.011$ at $1\,\sigma$ confidence level. Since the median value and the error bar of $n_s$ are smaller, when comparing with those from WMAP9 alone, the significance of $n_s < 1$ is more than $3\,\sigma$ confidence level. We also show the two-dimensional contour between $H_0$ and $n_s$ in the below panel of figure \ref{lcdm}. There is a strong correlation between $n_s$ and $H_0$, when considering the CMB data alone.

When combining all the datasets together, this strong degeneracy is broken apparently. The constraint on the Hubble constant becomes much more stringent, $h_0=0.6890\pm0.0062$ ($1\,\sigma$ C.L.). Consequently, the constraint of the spectral index also becomes tighter significantly, $n_s=0.963 \pm 0.008$ ($1\,\sigma$ C.L.). The error bar of $n_s$ is reduced by a factor of 1.5, due to the constraining power of BAO, HST and SN. The HZ spectrum is disfavored by the current data at about $5\,\sigma$ confidence level, which is consistent with that from the WMAP9 paper \cite{wmap9}.

However, this strong constraint on the spectral index is model-dependent. In some extended $\Lambda$CDM models, the constraints on $n_s$ could be changed, due to the possible degeneracies between $n_s$ and other extended parameters in different models. In the following sections, we discuss the constraints on parameters in these extended cosmological models, which is shown in table \ref{othertable}, such as the inflationary models, $\alpha_s$ and $r$, the dynamical dark energy model, $w(z)$, and ones including the neutrino properties, $\sum{m_\nu}$ and $N_{\rm eff}$.

\subsection{Inflationary Models}

\begin{figure}[t]
\begin{center}
\includegraphics[scale=0.45]{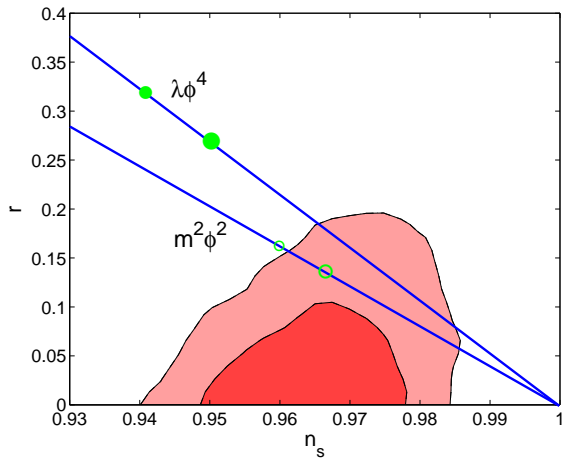}
\caption{Marginalized two-dimensional likelihood (1, $2\,\sigma$ contours) constraints on the parameters $n_s$ and $r$ from all datasets together (red). The two blue solid lines are predicted by the $m^2\phi^2$ and $\lambda\phi^4$ models, respectively. The green points denote predictions assuming that the number of e-foldings $N=50-60$ from two models.\label{lcdm_r}}
\end{center}
\end{figure}

\begin{table}
\caption{$1\,\sigma$ constraints on cosmological parameters $r$, $\alpha_s$, $\sum{m_\nu}$ and dark energy EoS from the current observations in different extended models. For the weakly constrained parameters we quote the $95\%$ upper limits instead.}\label{othertable}
\begin{center}

\begin{tabular}{c|c|c}

\hline\hline
models&constraints&$n_s$ constraints\\
\hline
$\Lambda$CDM+$r$&$r<0.15$&$0.966\pm0.009$\\
$\Lambda$CDM+$\alpha_s$&$\alpha_s=-0.023\pm0.011$&$0.948\pm0.011$\\
WCDM&$w=-1.060\pm0.066$&$0.960\pm0.009$\\
W(z)CDM&$w_0=-1.11\pm0.15$&$0.962\pm0.011$\\
&$w_1=0.18\pm0.65$&$-$\\
$\Lambda$CDM+$\sum{m_\nu}$&$\sum{m_\nu}<0.47$eV&$0.968\pm0.009$\\

\hline\hline
\end{tabular}
\end{center}
\end{table}

Firstly, we include the gravitational waves into the analysis. When using all datasets together, the data yield the $95\%$ upper limit of tensor-to-scalar ratio $r<0.15$. Meanwhile, the constraint on the spectral index is slightly relaxed, $n_s=0.966\pm0.009$ at $68\%$ confidence level, due to the degeneracy between $n_s$ and $r$. In figure \ref{lcdm_r} we show the two-dimensional constraints in the ($n_s$,$r$) panel which can be compared with the prediction of the inflation models. We find that the HZ scale-invariant spectrum ($n_s = 1$, $r = 0$) is still disfavored at about $4\,\sigma$ confidence level. Also, the inflation models with ``blue'' tilt ($n_s > 1$) are excluded by the current observations. Furthermore, assuming the number of e-foldings $N=50-60$, the single slow-rolling scalar field with potential $V(\phi)\sim m^2\phi^2$, which predicts ($n_s$,$r$)=($1-2/N$,$8/N$), is still within the $2\,\sigma$ region, while another single slow-rolling scalar field with potential $V(\phi)\sim \lambda\phi^4$, which predicts ($n_s$,$r$)=($1-3/N$,$16/N$), has been excluded more than at $2\,\sigma$ confidence level.

\begin{figure}[t]
\begin{center}
\includegraphics[scale=0.45]{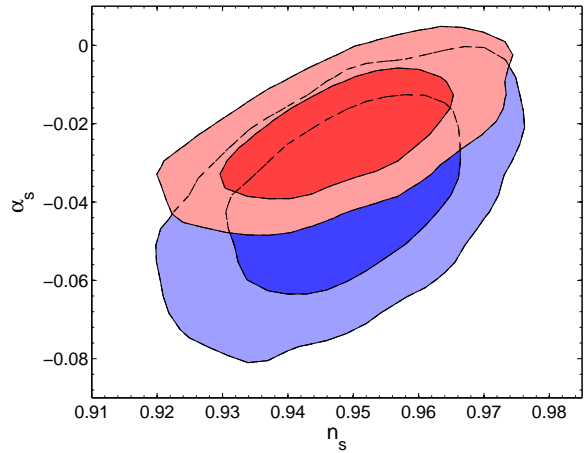}
\caption{Marginalized two-dimensional likelihood (1, $2\,\sigma$ contours) constraints on the parameters $n_s$ and $\alpha_s$ from all datasets together with (blue) and without (red) considering the tensor fluctuations in the analysis.\label{lcdm_as}}
\end{center}
\end{figure}

We also explore the constraint on the running of the spectral index from the latest observational data. When neglecting the tensor fluctuations ($r=0$), the combination of the current observational data yield the limit on the running of the spectral index of $\alpha_s=-0.023\pm0.011$ ($1\,\sigma$), which means the running of $n_s$ is favored by the current data at more than $2\,\sigma$ confidence level. In figure \ref{lcdm_as} we show the two-dimensional constraints in the ($n_s$,$\alpha_s$) panel. Due to the degeneracy between $n_s$ and $\alpha_s$, the $68\%$ constraint on $n_s$ is slightly enlarged, $n_s=0.948\pm0.011$. The error bar is relaxed by a factor of 1.5, when comparing with the standard $\Lambda$CDM model.

Finally, we vary the $\alpha_s$ and $r$ simultaneously in the analysis. From the blue contour of figure \ref{lcdm_as}, one can see that the constraint on $n_s$ does not change, $n_s=0.949\pm0.011$ ($1\,\sigma$). The degeneracy between $\alpha_s$ and $r$ significantly weakens the constraints on them, namely the $68\%$ constraint on $\alpha_s$ is $\alpha_s=-0.039\pm0.016$ and the $95\%$ upper limit on $r$ is $r < 0.35$. The current data still favor the running of $n_s$ at more than $2\,\sigma$ confidence level.

\begin{figure}[t]
\begin{center}
\includegraphics[scale=0.45]{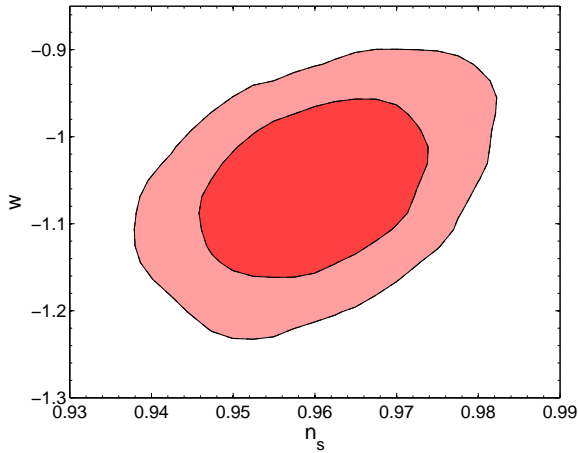}
\caption{Marginalized two-dimensional likelihood (1, $2\,\sigma$ contours) constraints on the parameters $n_s$ and $w$ from all datasets together (red).\label{lcdm_w}}
\end{center}
\end{figure}

\subsection{Dynamical Dark Energy}

\begin{figure}[t]
\begin{center}
\includegraphics[scale=0.45]{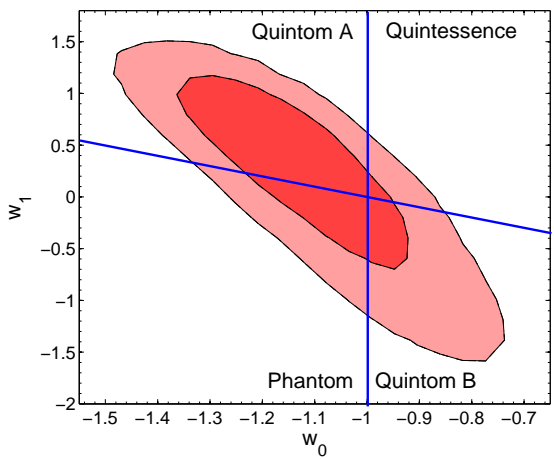}
\caption{Marginalized two-dimensional likelihood (1, $2\,\sigma$ contours) constraints on the parameters $w_0$ and $w_1$ from all datasets together (red). The blue solid lines stand for $w_0 =-1$ and $w_0 + w_1 =-1$.\label{lcdm_wz}}
\end{center}
\end{figure}

Assuming the flat universe, first we explore the cosmological constraints in the dark energy model with a constant EoS, $w$ ($w \equiv w_0$, $w_1 \equiv 0$), from the latest observational data. In figure \ref{lcdm_w} we show the two-dimensional constraints on $w$ and $n_s$. Current observational data yield a strong constraint on the constant EoS of dark energy, $w=-1.060\pm0.066$ ($1\,\sigma$), which is similar with the limit from WMAP9 \cite{wmap9}. The standard $\Lambda$CDM model ($w=-1$) is consistent with the current observational data. In this case the constraint on $n_s$ is slightly changed, $n_s=0.960\pm0.009$ at $68\%$ confidence level, due to the degeneracy between $n_s$ and $w$.

For the time evolving EoS, in figure \ref{lcdm_wz} we illustrate the constraints on the dark energy parameters $w_0$ and $w_1$. For the flat universe, due to the limits of the precisions of observational data, the variance of $w_0$ and $w_1$ are still large, namely, the $68\%$ constraints on $w_0$ and $w_1$ are $w_0=-1.11\pm0.15$ and $w_1=0.18\pm0.65$. And the $95\%$ constraints are $-1.38<w_0<-0.80$ and $-1.32<w_1<1.15$. This result implies that the dynamical dark energy models are not excluded and the current data cannot distinguish different dark energy models decisively. The obtained best fit model is the Quintom dark energy model \cite{quintom} with the particular feature that its EoS can cross the cosmological constant boundary smoothly. The standard $\Lambda$CDM model, however, is still a good fit right now.

For the spectral index, the weak correlations between $n_s$ and the parameter of dark energy EoS do not change the limit significantly, namely the $68\%$ constraint is $n_s=0.962\pm0.011$. The HZ spectrum is still ruled out at about $4\,\sigma$ confidence level.

\subsection{Neutrino Properties}

\begin{figure}[t]
\begin{center}
\includegraphics[scale=0.45]{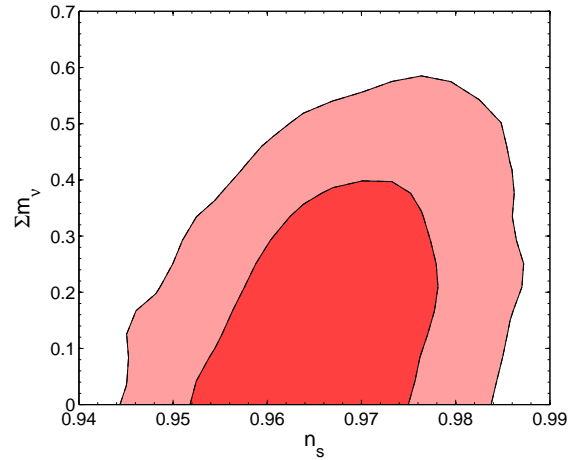}
\caption{Marginalized two-dimensional likelihood (1, $2\,\sigma$ contours) constraints on the parameters $n_s$ and $\sum{m_\nu}$ from all datasets together (red).\label{lcdm_nu}}
\end{center}
\end{figure}

\begin{figure}[t]
\begin{center}
\includegraphics[scale=0.45]{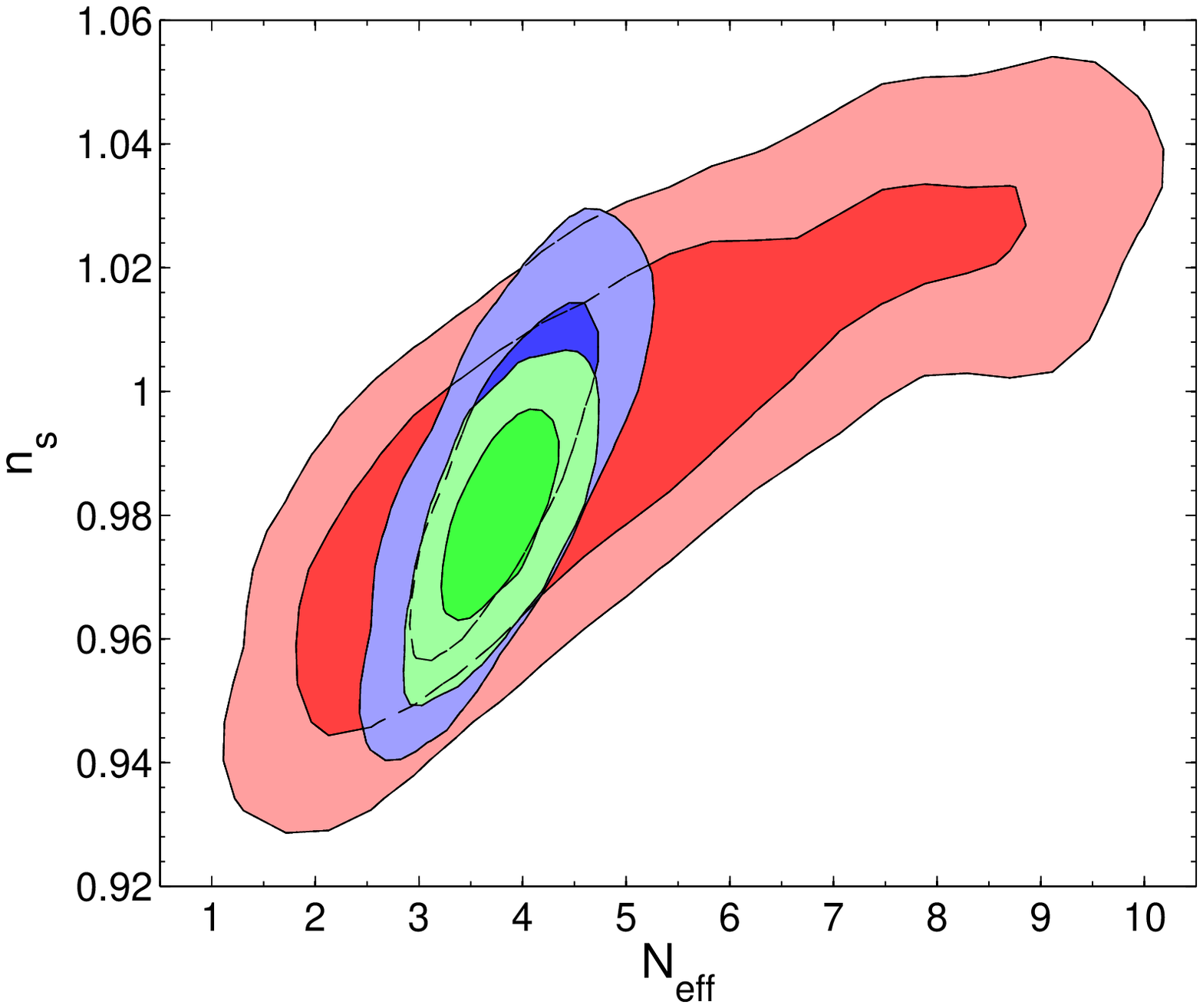}
\caption{Marginalized two-dimensional likelihood (1, $2\,\sigma$ contours) constraints on the parameters $n_s$ and $N_{\rm eff}$ from different present data combinations: WMAP9 only (red), WMAP9+SPT (blue) and All datasets (green).\label{lcdm_neff}}
\end{center}
\end{figure}

\begin{table}
\caption{$1\,\sigma$ constraints on some cosmological parameters from different data combinations in the model with the number of relativistic species $N_{\rm eff}$.}\label{nefftable}
\begin{center}

\begin{tabular}{c|c|c|c}

\hline\hline
\multicolumn{4}{c}{$\Lambda$CDM + $N_{\rm eff}$}\\
\hline
&WMAP9 alone&WMAP9+SPT&All Datasets\\
\hline
$n_s$&$0.994\pm0.026$&$0.986\pm0.018$&$0.980\pm0.011$\\
$N_{\rm eff}$&$4.88\pm2.08$&$3.82\pm0.57$&$3.79\pm0.38$\\
$H_0$&$76.84\pm8.18$&$74.31\pm3.60$&$72.29\pm1.82$\\
$100\Omega_bh^2$&$2.259\pm0.050$&$2.271\pm0.048$&$2.252\pm0.036$\\
$100\Omega_ch^2$&$14.75\pm3.79$&$12.68\pm1.04$&$13.04\pm0.73$\\

\hline\hline
\end{tabular}
\end{center}
\end{table}

Finally, we study the constraint on $n_s$ in the cosmological models including the neutrino properties, the massive neutrino and the number of relativistic species. In table \ref{othertable} we show the constraint on the total neutrino mass from all datasets together, namely the $95\%$ upper limit is $\sum{m_\nu}<0.47$eV, which is consistent with previous works \cite{wmap9,Xia:2012na,Zhao:2012xw,viel,srs}. In figure \ref{lcdm_nu} we show the constraints in the ($n_s$,$\sum{m_\nu}$) panel. Since the degeneracy between $n_s$ and $\sum{m_\nu}$ is not very strong, including the massive neutrino does not change the constraint on $n_s$ significantly, namely $n_s=0.968\pm0.009$ at $68\%$ confidence level.

Then we consider the constraints on the number of relativistic species, $N_{\rm eff}$ , from different data combinations (Table \ref{nefftable}), assuming massless neutrinos. We find the WMAP9 data alone gives very weak constraint $N_{\rm eff}=4.88\pm2.08$ at the $68\%$ confidence level, consistent with the result derived by the WMAP9 team. Adding the small-scale SPT data significantly improves the constraints on $N_{\rm eff}$ to $N_{\rm eff}=3.82\pm0.57$ ($1\,\sigma$). When we combine all datasets together, we obtain our most stringent constraint of $N_{\rm eff}=3.79\pm0.38$ ($68\%$ C.L.). Our results are quite consistent with previous works \cite{Xia:2012na,Zhao:2012xw,spt}, and display a slight preference for an extra relativistic relic. However, the standard value of $N_{\rm eff}=3.04$ remains well within the $95\%$ confidence intervals.

Since $N_{\rm eff}$ can be written in terms of $\Omega_mh^2$ and the redshift of matter-radiation equality, $z_{\rm eq}$, there are strong degeneracies present between $N_{\rm eff}$, the matter density,
$\Omega_mh^2$ and the Hubble parameter $H_0$. Consequently, $N_{\rm eff}$ is also strong correlated with the spectral index $n_s$. In figure \ref{lcdm_neff} we show the constraints on $n_s$ and $N_{\rm eff}$ from different data combinations. When using all datasets together, the $68\%$ C.L. constraint on $n_s$ becomes $n_s=0.980\pm0.011$. The HZ spectrum now is consistent with the current data at $95\%$ confidence level.

When we include the number of relativistic species, $N_{\rm eff}$, and the massive neutrino simultaneously, the constraints on parameters become weaker, due to the degeneracies among them. The $95\%$ upper limit of the total neutrino mass is $\sum{m_\nu}<0.71$. The $68\%$ C.L. constraints on $N_{\rm eff}$ and $n_s$ becomes $N_{\rm eff}=3.89\pm0.39$ and $n_s=0.988\pm0.013$.


\section{Summary}\label{summary}

Recently many experimental groups have published their new observational data, such as temperature and polarization power spectra of WMAP9 \cite{wmap9}, temperature power spectrum of SPT at high multipoles $\ell$ \cite{spt}, and the BAO measurement from the Ly-$\alpha$ forest of BOSS quasars at high redshift $z=2.3$ \cite{highbao}. The WMAP collaboration has presented the cosmological implications of their final nine-year data release, finding that the spectra with the spectral index $n_s \geq 1$ are disfavored by the current observational data at about $5\,\sigma$ confidence level in the $\Lambda$CDM framework.

However, in the analyses we find that the strong constraint on $n_s$ could be weakened by considering the possible degeneracies between $n_s$ and other cosmological parameters introduced in some extended models, such as the tensor fluctuation $r$ and the dark energy EoS $w$. The largest effect is shown in the model with the number of relativistic species $N_{\rm eff}$. Due to the strong degeneracy between $n_s$ and $N_{\rm eff}$, the error bar of $n_s$ is significantly enlarged, namely $n_s=0.980\pm0.011$ ($1\,\sigma$), and the HZ spectrum now is consistent with the current data at $95\%$ confidence level.

Finally, we vary all of the extended cosmological parameters at the same time. Due to the degeneracies among these parameters, their constraints become much weaker. More importantly, the limit on the spectral index $n_s$ is relaxed further, $n_s=0.977\pm 0.024$ at $68\%$ confidence level. The disagreement between the HZ spectrum and the current observational data disappears. The high accurate Planck measurement \cite{planck} is needed, in order to constrain $n_s$ and distinguish various inflationary models better.


\section*{Acknowledgements}

We acknowledge the use of the Legacy Archive for Microwave Background Data Analysis (LAMBDA). Support for LAMBDA is provided by the NASA Office of Space Science. HL is supported in part by the National Science Foundation of China under Grant Nos. 11033005, by the 973 program under Grant No. 2010CB83300, by the Chinese Academy of Science under Grant No. KJCX2-EW-W01. JX is supported by the National Youth Thousand Talents Program and the grants No. Y25155E0U1 and No. Y3291740S3. XZ is supported in part by the National Science Foundation of China under Grants No. 10975142 and 11033005, and by the Chinese Academy of Sciences under Grant No. KJCX3-SYW-N2.


\end{document}